\documentclass[12pt, a4paper]{article}
\usepackage{geometry}
\usepackage[utf8]{inputenc}

\usepackage[T1]{fontenc} % if needed

\usepackage{bm}
\usepackage{secdot}
\usepackage{mathptmx}
\usepackage{float}
\usepackage{soul}
\usepackage{authblk}

\RequirePackage{graphicx}
\RequirePackage{amsmath}
\RequirePackage{amssymb}

\title{Reconstruction of $f(R)$ Gravity from Cosmological Unified Dark Fluid Model} 

\author{Esraa Ali Elkhateeb\footnote{dr.esraali@sci.asu.edu.eg}}

\affil{Ain Shams University, Faculty of Science, Physics Department, Abbaseia, 11566, Cairo, Egypt.}

\begin{document}

\maketitle

\abstract{In this work, we reconstruct the cosmological unified dark fluid model proposed previously by Elkhateeb \cite{Elkhateeb:2017oqy} in the framework of $f(R)$ gravity. Utilizing the equivalence between the scalar-tensor theory and the $f(R)$ gravity theory, the scalar field for the dark fluid is obtained, whence the $f(R)$ function is extracted and its viability is discussed. The $f(R)$ functions and the scalar field potentials have then been extracted in the early and late times of asymptotically de Sitter spacetime. The ability of our function to describe early time inflation is also tested. The early time scalar field potential is used to derive the slow roll inflation parameters. Our results of the tensor-to-scalar ratio $r$ and the scalar spectral index $n_s$ are in good agreement with results from Planck-2018 TT+TE+EE+lowE data for the model parameter $n > 2$.
}
%%%%%%%%%%%%%%%%%%
%\keywords{Cosmology, cosmic inflation, cosmological evolution, dynamical evolution}
%Dark matter,  dark energy, early Universe, inflation, observations.}
%%%%%%%%%%%%%%%%%%%%%%%%%%%%%%%%%
\section{Introduction}
Unexpected late time accelerating expansion of the universe is interpreted in the framework of general relativity by the presence of a peculiar cosmic fluid with negative pressure, the dark energy. The nature of dark energy is still a mystery and remains an active area of research. The simplest explanation was the cosmological constant, leading to the $\Lambda$CDM model. Although being the best fit for most of the data, the cosmological constant approach suffers many problems\cite{Carroll:2000fy,Lombriser:2019jia,Wondrak:2017eao}. An example is the lack of physical origin. The cosmological constant was introduced theoretically by Einstein in 1917 in his gravitational field equations \cite{Einstein:1917x}. As a constant, this term is interpreted as the intrinsic energy density of the vacuum, and in turn, finds its origin in the quantum field theory (QFT) of particles and fields. QFT, on the other hand, predicts a vacuum energy density (VED) disastrously larger than the current observed critical density of the universe, from which the VED is one part. Unfortunately, models tried to explain the value of the observed VED all failed, and an obstacle is placed in the way, leading Weinberg to formulate his "no-go theorem" \cite{Weinberg:1988cp}. Another problem is the present-time coincidence of the measured order of magnitude of the VED and the matter-energy density, leading to the famous cosmological coincidence problem. Such problems have motivated a host of other alternatives. Dynamical energy models associated with scalar fields, quintessence, are one of the most studied candidates for dark energy \cite{Tsujikawa:2013fta,Caldwell:2000wt}.

On the other hand, the dark side of the universe contains another mysterious element, dark matter, a peculiar matter which is non-baryonic and interacts gravitationally. The unknown nature of dark energy and dark matter motivated the idea of the unification of these two dark sectors. Models that adopted this idea introduced a single fluid, dubbed dark fluid, with an equation of state that enables a smooth transition from the dark matter domination era to the dark energy domination era. Chaplygin gas model and its generalizations are examples of these models \cite{Kamenshchik:2001cp,Bento:2002ps}. Another example is the new unified dark fluid model (UDF) proposed by the present author \cite{Elkhateeb:2017oqy}.

The above models aim to modify the energy-momentum tensor on the right-hand side of the Einstein field equations. Another line of thought assumes modification on the left-hand side of the equations. In this case, cosmic acceleration is considered to be a consequence of gravitational physics. The class of models adopted in this approach is known as modified gravity models \cite{Shankaranarayanan:2022wbx,Sebastiani:2016ras,Nojiri:2006ri,Carroll:2003wy,Faraoni:2005vk,Abdalla:2004sw,Li:2007xn,Capozziello:2005ku,Starobinsky:1980te,Copeland:2006wr}. One of the extensively studied classes of this approach is the $f(R)$ gravity models \cite{Hu:2007nk,Appleby:2007vb,Amendola:2006we,DeFelice:2010aj,delaCruz-Dombriz:2006kob,Cognola:2005de,Starobinsky:2007hu}. In these models, an arbitrary function $f(R)$ replaces the Ricci scalar $R$ in the Einstein-Hilbert action. In that sense, Einstein's general relativity is recovered by considering $f(R)=R$. Some modified $f(R)$ gravities have been proposed to describe the late time acceleration of the universe \cite{Appleby:2007vb,Amendola:2007nt}. Others can unify the late acceleration and the early inflation and grant a successful transition from the matter domination phase to the late time acceleration phase of the universe \cite{Granda:2020dty,Nojiri:2019fft,Nojiri:2010ny,Nojiri:2010wj,Nojiri:2007as}. Moreover, modified gravity can play the role of dark matter \cite{Vagnozzi:2017ilo,Myrzakulov:2015kda,Cembranos:2015svp}. Cosmological observations and solar system tests can constrain such models. Reconstruction schemes for modified gravity theories in general and the $f(R)$ gravity, in particular, can be found in \cite{Gadbail:2023klq,Gadbail:2022jco,Odintsov:2020fxb,Odintsov:2017fnc,Nojiri:2009kx,Frolov:2008uf,Nojiri:2006gh,Nojiri:2006be}. A comprehensive review on modified gravity theories covering the latest developments in the subject can be found in \cite{Nojiri:2017ncd}.

This work aims to reconstruct the new cosmological unified dark fluid model, UDF, in the framework of $f(R)$ gravity. The UDF model was suggested by the present author in 2018 \cite{Elkhateeb:2017oqy}. It introduces a single unified dark fluid with a particular equation of state that enables the fluid to smoothly evolve from the matter era to the dark energy era. The work was extended by adding viscosity to the fluid in 2019 \cite{Elkhateeb:2018unf}. Recently, the model was tested against observational data from local $H_0$ measurements, type Ia supernovae, observational Hubble data, baryon acoustic oscillations, and cosmic microwave background \cite{Elkhateeb:2021atj}. The Akaike Information Criterion (AIC) results revealed that this UDF model is substantial and is favored by the observational data used over the $\Lambda$CDM model.

The paper is organized as follows: in Sec. 2 we present a brief description of the scalar-tensor theory and its connection to the $f(R)$ gravity. In Sec. 3 we review the new cosmological UDF model under consideration. In Sec. 4 we reconstruct this UDF model in the framework of $f(R)$ gravity and discuss the function viability. In Sec. 5 we study the inflationary behavior of the function and compare our results with the recent data from Planck 2018. In the last Section, we make a brief conclusion.
%%%%%%%%%%%%%%%%%%%%%%%%%%%%%%%%%%%%%%%%%%%%%%%%%%%%%%%%%%%%%

\section{Metric $f(R)$ Gravity vs Scalar Tensor Theories}
One of the simplest models of modified gravity is the $f(R)$ gravity.  The general 4-dimensional action is 

\begin{equation} \label{action}
    S_{grav}=\frac{1}{2 \kappa^2} \int d^4 x\; \sqrt{-g}\; f(R)\; +\; S_m\left(g_{\mu \nu}, \Psi_m\right)
\end{equation}
Where $\kappa^2=8 \pi G$, $S_m$ is the matter action and $\Psi_m$ is the matter fields. By varying the action (\ref{action}) with respect to $g_{\mu \nu}$ the field equations can be obtained, namely,

\begin{equation} \label{fieldeq}
   R_{\mu \nu}\; f'(R)\; - \frac{1}{2} g_{\mu \nu} f(R)\; - \nabla_\mu \nabla_\nu f'(R)\; + g_{\mu \nu} \square f'(R)\;=\;T_{\mu \nu} 
\end{equation}
Where $T_{\mu \nu}$ is the matter energy-momentum tensor. Considering units where $8 \pi G\;=\;1$, the contraction of eq. (\ref{fieldeq}) gives

\begin{equation} \label{contr}
   3\; \square f'(R)\; +\; R\; f'(R)\; -\; 2\; f(R) \;=\; \rho\; -\; 3\; P
\end{equation}
Where we used the relation
\begin{equation} \label{Tvalue}
   T\;=\; g^{\mu \nu}\; T_{\mu \nu}\;= \rho\; -\; 3\; P 
\end{equation}
and $\rho$ and $P$ are the energy density and pressure of matter, respectively.

The non-vanishing $\square f'(R)$ term in eq.(\ref{contr}) indicates that there exists a propagating scalar degree of freedom, named "scalaron", whose dynamics are governed by this equation.  For vacuum solution, the $\square f'(R)$ term vanishes, while Ricci scalar turns to be constant, and we get
\begin{equation} \label{vacuum}
    \; R\; f'(R)\; -\; 2\; f(R) \;=\; 0
\end{equation}

The occurrence of the scalaron field realizes the idea of Branse-Dicke (BD) theory, where a scalar field mediates gravity along with the metric tensor field \cite{Brans:1961sx,Polarski:2000ps,Fujii:2003pa}. To establish this connection, let's start with the $f(R)$ gravity action (\ref{action}). One can introduce a scalar field $\chi$ with the scalar-tensor action integral \cite{Teyssandier:1983zz}
\begin{equation} \label{STact}
    S_{ST}=\frac{1}{2 \kappa^2} \int d^4 x\; \sqrt{-g}\; \left[f(\chi) + \left(R\; -\; \chi\right)\; f'(\chi)\right]\; +\; S_m\left(g_{\mu \nu}, \Psi_m\right)
\end{equation}
Variation of this action with respect to the scalar field $\chi$ yields the relation
\begin{equation} \label{chiR}
    \left(R\; -\; \chi\right)\;f''\; =\; 0
\end{equation}
As a consequence, $\chi\; =\; R$ provided that $f''\; \ne \; 0$.

Now let's redefine the field $\chi$ through another fields $\Phi$ and $\varphi$ such that
\begin{equation} \label{pPhid}
    \Phi \;=\; f'(\chi)
\end{equation}
and \cite{Frolov:2008uf}
\begin{equation} \label{pvarphid}
    \varphi \;=\; f'(\chi) \;-\; 1 
\end{equation}
So that $\Phi=\varphi+1$. Accordingly, action (\ref{STact})  reduces for the filed $\Phi$ to
\begin{equation} \label{BDact}
    S_{ST}=\frac{1}{2 \kappa^2} \int d^4 x\; \sqrt{-g}\; \left(\Phi \; R\; - \; V(\Phi)\right)\; +\; S_m\left(g_{\mu \nu}, \Psi_m\right)
\end{equation}
where
\begin{equation} \label{Vst}
    V(\Phi) \;=\; \Phi \; \chi \; -\; f\left(\chi(\Phi)\right)
\end{equation}
Notice that the same action is obtained for the field $\varphi$ due to eq.(\ref{chiR}). 

Recalling that the BD action has the form
\begin{equation} \label{BDactG}
    S_{BD}=\frac{1}{2 \kappa^2} \int d^4 x\; \sqrt{-g}\; \left(\Phi \; R\; -\; \frac{\omega}{\Phi} \nabla_{\mu} \phi \nabla^\mu \Phi \;- \; V(\Phi)\right)\; +\; S_m\left(g_{\mu \nu}, \Psi_m\right)
\end{equation}
where $\omega$ is a constant known as the BD parameter. One can see that action (\ref{BDact}) represents the BD action for $\omega=0$. Variation of action (\ref{BDact}) with respect to the metric results in the field equation for the field $\Phi$, namely
\begin{equation} \label{fieldphi}
   G_{\mu \nu}\; \Phi\; - \nabla_\mu \nabla_\nu \Phi\; + g_{\mu \nu} \square \Phi\; + \frac{1}{2} g_{\mu \nu} V(\Phi)\; =\;T_{\mu \nu} 
\end{equation}
Noting that the field equation for $f(R)$ gravity, eq.(\ref{fieldeq}), can be rewritten as
\begin{equation} \label{fieldfR}
   G_{\mu \nu}\; f'(R)\; - \frac{1}{2} g_{\mu \nu} f(R)\; + \frac{1}{2} g_{\mu \nu} \;R\; f'(R)\; - \nabla_\mu \nabla_\nu f'(R)\; + g_{\mu \nu} \square f'(R)\;=\;T_{\mu \nu} 
\end{equation}
Equations (\ref{fieldphi}) and (\ref{fieldfR}) for the field equations of BD theory and $f(R)$ gravity show a one-to-one correspondence provided that
\begin{equation} \label{phid}
    f'(R) = \Phi = \varphi-1 .
\end{equation}

%%%%%%%%%%%%%%%%%%%%%%%%%%%%%%%%%%%%%%%%%%%%%%%%%%%%%%%%%%%%%

\section{The Cosmological UDF Model}
The UDF model is a barotropic unified dark fluid characterized by the equation of state (EoS)
\begin{equation} \label{eos}
    P\;=\; -\rho \;+\; \frac{\alpha \; \rho^m}{1 \;+\; \beta\; \rho^n}.
\end{equation}

The two asymptotes of this EoS are power laws for the density. In consequence, the fluid evolves smoothly from dust at early times to dark energy at late times. One can constrain some of the model parameters by studying the asymptotic behavior of its EoS. Following \cite{Elkhateeb:2021atj}, we get
%Studying the asymptotic behavior of the model can shed light on the relationship between its parameters. Following \cite{Elkhateeb:2021atj}, we get
\begin{equation} \label{params}
    m \;=\; n \;+\; 1\; ;  \;\;\;\;\;\;\;\;\;\;\;\;\;\;   \alpha \;=\; \beta
\end{equation}
Defining
\begin{equation} \label{def}
    \Tilde{\beta}  \;=\; \beta^{-\frac{1}{n}}
\end{equation}
which has the dimensions of $H^2$; $\left[\Tilde{\beta}\right] = \left[H^2\right]$. The EoS of the UDF model can then be written as
\begin{equation} \label{eos1}
    P\;=\; -\rho \;+\; \frac{\rho^{n+1}}{\Tilde{\beta}^n \;+\; \rho^n}
\end{equation}
Using this in the conservation equation for the energy density
\begin{equation}  \label{cons}
    \frac{d \rho}{d t}\;=\; -3\; \frac{\Dot{a}}{a} \left(\rho \;+\; P\right)
\end{equation}
we get the relation governing the evolution of the energy density as a function of the scale factor,
\begin{equation} \label{rho}
    \rho(a) = \rho_0 \left(\frac{a}{a_0}\right)^{-3} e^{-\frac{\Tilde{\beta}^n}{n \rho_0^n}} \; {\rm exp}\left\{\frac{1}{n} W\left[\left(\frac{a}{a_0}\right)^{3n} \left(\frac{\Tilde{\beta}}{\rho_0}\right)^n e^{\frac{\Tilde{\beta}^n}{\rho_0^n}}\right] \right\}
\end{equation}
Where $W$ is the Lambert $W$ function. The parameters $n$ and $\Tilde{\beta}$ can be constrained using observational data.

On the other hand, the energy density and pressure for the scalar field are given by 
\begin{equation} \label{rho-phi}
    \rho_\varphi = \frac{1}{2} \Dot{\varphi}^2 \;+\; V(\varphi)
\end{equation}
\begin{equation} \label{p-phi}
    P_\varphi = \frac{1}{2} \Dot{\varphi}^2 \;-\; V(\varphi)
\end{equation}
Adding (\ref{rho-phi}) and (\ref{p-phi}), and using eqn.(\ref{eos1}) we get
\begin{equation} \label{phi}
   \Dot{\varphi}^2 = \rho \;+\; P \;=\; \frac{\rho^{n+1}}{\Tilde{\beta}^n \;+\; \rho^n}
\end{equation}
Now using
\begin{equation} \label{dif}
    \Dot{\varphi} \;=\; \frac{d\varphi}{d\rho} \frac{d\rho}{dt} 
\end{equation}
together with (\ref{cons}) and (\ref{phi}), we get
\begin{equation} \label{dphidrho}
    \frac{d\varphi}{d\rho} \;=\; \mp \frac{1}{\sqrt{3}} \left(\frac{\Tilde{\beta}^n \;+\; \rho^n}{\rho^{n+2}}\right)^\frac{1}{2}
\end{equation}
This has the solution
\begin{equation} \label{phi1}
    \varphi(\Tilde{\rho}) \;=\; \pm \frac{2}{n \sqrt{3}} \left({\rm arcsinh}{\left(\sqrt{\Tilde{\rho}^n}\right)}-\sqrt{\frac{\Tilde{\rho}^n+1}{\Tilde{\rho}^n}}\right) \;+\; C
\end{equation}
Where $\Tilde{\rho}\;=\; \rho/\Tilde{\beta}$ is dimensionless, and $C$ is the integration constant.

The potential $V$, on the other hand, is given from (\ref{rho-phi}) and (\ref{p-phi}) to be
\begin{eqnarray}
   V \;=\; \frac{1}{2} \left(\rho \;-\; P\right) \;=\;  \frac{1}{2} \left(2\;\rho \;-\; \frac{\rho^{n+1}}{\Tilde{\beta}^n \;+\; \rho^n}\right)  \label{V1} \\
   \;=\;  \frac{1}{2} \Tilde{\beta}\;\Tilde{\rho} \left(\frac{2\;+\;\Tilde{\rho}^{n}}{1 \;+\; \Tilde{\rho}^n}\right)    \label{V}
\end{eqnarray}
Where we used eqn.(\ref{eos1}).

%%%%%%%%%%%%%%%%%%%%%%%%%%%%%%%%%%%%%%%%%%%%%%%%%%%%%%%%%%%%%

\section{Reconstruction of $f(R)$ gravity from the UDF model}
The generalized Friedmann equations in the $f(R)$ gravity can be obtained from the time-time and space-space components of the field equation (\ref{fieldeq}). In the flat FLRW spacetime these reads
\begin{equation} \label{fried1}
    \left(\frac{\Dot{a}}{a}\right)^2 \;=\; \frac{1}{3\;F} \left[\rho\; - \frac{1}{2} \left(f-R\; F\right)\; -3\;H\;\Dot{F}\right]
\end{equation}
\begin{equation} \label{fried2}
    \left(2\;\frac{\Ddot{a}}{a} \;+\;H^2\right) \;=\; \frac{-1}{2\;F} \left[2\; \Ddot{F} \;+\; 4\;H\;\Dot{F} \; +\; f \;-\; R\; F \right] \;-\;\frac{1}{F}\;P
\end{equation}
with $F(R)=f'(R)$. While the Ricci scalar is given by
\begin{equation} \label{R-H}
    R \;=\; 6\left(2\;H^2 \;+\; \Dot{H} \right)
\end{equation}
We can then define a curvature density and a curvature pressure as
\begin{eqnarray}  \label{crhoP}
  \begin{split}
    \rho_{curv} & = \left(\frac{R\; F -f-6\;H\;\Dot{F}}{2\;F} \right) \\
    P_{curv} & = \left(\frac{2\; \Ddot{F} \;+\; 4\;H\;\Dot{F} \; +\; f \;-\; R\; F}{2\;F} \right)
 \end{split}
\end{eqnarray} 
The generalized Friedmann equations (\ref{fried1}) and (\ref{fried2}) now take the form
\begin{equation} \label{fried3}
    \begin{split}
       H^2&=\frac{1}{3} \; \rho_{eff}  \\
    \frac{\Ddot{a}}{a}&=\frac{-1}{6} \left(\rho_{eff} \;+\; 3\;P_{eff}\right) 
    \end{split}
\end{equation}
Where
\begin{equation} \label{eff}
    \rho_{eff} \;=\; \Bar{\rho} \;+\; \rho_{curv}  \;\;\;\;\;\;\;\; \& \;\;\;\;\;\;\;\; p_{eff} \;=\; \Bar{P} \;+\; P_{curv}
\end{equation}
with $\Bar{\rho}$ and $\Bar{P}$ are given by the regular matter-components density and pressure, $\rho$ and $P$, divided by $F$. The Ricci scalar can now be written as
\begin{eqnarray}  \label{par}
  \begin{split}
    R & = \rho_{eff} \;-\; 3\;p_{eff} \\
    & = (1\;-\;3\;\omega_{eff}) \; \rho_{eff} \;=\; \gamma \; \rho_{eff}
 \end{split}
\end{eqnarray}
where $\omega_{eff}$ is the equation of state parameter, and $\gamma$ is defined by $(1\;-\;3\;\omega_{eff})$. Neglecting the contribution of any other kind of matter, we can identify $\rho_{curv}$ and $ P_{curv}$ in eq.(\ref{crhoP}) with the density and pressure of the unified dark fluid UDF. The equation of state parameter can then be found from eq.(\ref{eos1}) to be
\begin{equation}
    \omega_{eff} = -1\ \;+\; \frac{\rho^{n}}{\Tilde{\beta}^n \;+\; \rho^n} 
\end{equation}
Using this in (\ref{par}), we get
\begin{eqnarray}
   R \;=\; \; 4\; \rho \;-\; \frac{3\;\rho^{n+1}}{\Tilde{\beta}^n \;+\; \rho^n}    \label{R1} \\
   \;=\; \Tilde{\beta}\;\Tilde{\rho} \left(\frac{4\;+\;\Tilde{\rho}^{n}}{1 \;+\; \Tilde{\rho}^n}\right)    \label{R2}
\end{eqnarray}
Which is rearranged to give
\begin{equation} \label{difeq}
    \left(1 \;+\; \Tilde{\rho}^n\right)\; R \;-\; 4\; \Tilde{\beta}\;\Tilde{\rho} \;-\; \Tilde{\beta}\;\Tilde{\rho}^{n+1} \;=\;0
\end{equation}

The field function $\varphi$ can then written from (\ref{phi1}) as
\begin{equation} \label{philte}
    \varphi(R) \;=\; \pm \frac{2}{n \sqrt{3}} \left[{\rm arcsinh} \left({\sqrt{\left( \frac{R}{\gamma\; \Tilde{\beta}} \right)^n}} \right)
    -\sqrt{1 + \left( \frac{R}{\gamma\; \Tilde{\beta}} \right)^{-n}}\right] \;+\; C
\end{equation}
Using (\ref{phid}), we then have
\begin{equation}
  \begin{split}
     f(R) & = R \pm \frac{2}{n \sqrt{3}}  \left\{  \left({\rm arcsinh} \sqrt{\left(\frac{R}{\gamma \Tilde{\beta}}\right)^n}\right) R \right.  
      - {_2}F_1\left(-\frac{1}{2},-\frac{1}{n};1-\frac{1}{n};-\left(\frac{R}{\gamma \Tilde{\beta}}\right)^{-n}\right) \; R  \\
     & \hspace{2.2cm} \left. - \frac{n\; R}{(n+2)} \; \sqrt{\left(\frac{R}{\gamma \Tilde{\beta}}\right)^{n}} \; {_2}F_1\left(\frac{1}{2},\frac{1}{2}+\frac{1}{n};\frac{3}{2}+\frac{1}{n};-\left(\frac{R}{\gamma \Tilde{\beta}}\right)^n \right) \right\} \;+\; C_1
  \end{split}
\end{equation}
where the function $_2F_1(a,b;c;z)$ is the Gaussian hypergeometric function.

The viability of $f(R)$ demands that $f'(R) > 0$ and $f''(R) > 0$ for $R \ge R_0$ where $R_0$ is the Ricci scalar at the present epoch \cite{Faraoni:2005vk,Amendola:2006we,DeFelice:2010aj,Amendola:2007nt,Olmo:2005hc}.  These conditions guarantee stability under perturbation, ensure keeping sign of the effective Newtonian coupling, and avoid the negativity of the squared mass of the scalaron field. Testing our $f (R)$ we find that while $f'(R) > 0$ for both $\pm$ sign functions, $f''(R) < 0$ for the $-ve$ sign one, meaning that the $-ve$ sign function is not viable. Accordingly, only the function with the $+ve$ sign will be considered. In the following, We study the behavior of our $f(R)$ function in the early and late universe.
%%%%%%%%%%%%%%%%%%%%%%%%%%%%%%%%%%

\subsection{Late Universe}
For late universe, where $\Tilde{\rho}$ is small, relation (\ref{phi1}) gives
\begin{equation} \label{latephi}
    \varphi(\Tilde{\rho}) \;\approx\; -\; \frac{2}{n \sqrt{3}} \frac{1}{\sqrt{\Tilde{\rho}^n}}
\end{equation}
So that 
\begin{equation}
    \Tilde{\rho} \;\approx\; \frac{4}{3\;n^2}\; \varphi^{-2/n}
\end{equation}
On the other hand, we have from (\ref{R2})
\begin{equation} \label{lateR}
    R \;\approx\; 4\; \Tilde{\beta}\;\Tilde{\rho}
\end{equation}
So that
\begin{equation} \label{latephiR}
    \varphi(R) \;\approx\; -\; \frac{2}{n \sqrt{3}} \frac{1}{\sqrt{R^n/(4\;\Tilde{\beta})^n}}
\end{equation}
While treating $V(\Tilde{\rho})$ in (\ref{V}) for small $\Tilde{\rho}$, we get
\begin{equation}   \label{lateV}
    V(\Tilde{\rho}) \approx \Tilde{\beta}\;\Tilde{\rho}
\end{equation}
So that
\begin{equation} \label{lateVR}
    V(\varphi) \;\approx\; -\frac{4\;\Tilde{\beta}}{3\;n^2}\; \varphi^{-2/n}
\end{equation}
Using eq. (\ref{latephiR}) in (\ref{phid}) and integrate, we get

\begin{equation} \label{latef}
    f(R) \;=\;R \;+\; \frac{4\;(4\;\Tilde{\beta})^{n/2}}{n\;(n-2)\;\sqrt{3}}\; R^{(1-n/2)}
\end{equation}
This is a common form for $f(R)$. For $n < 2$ this function takes the general form
\begin{equation}
    f(R) = R \; - \; \alpha \; R^k \; ; \;\;\;  k < 1 , \alpha > 0
\end{equation}
This form of $f(R)$ was studied in detail in \cite{Amendola:2006we}. It was found that this model gives rise to a damped oscillation around the matter era and can reach a stable de Sitter point.

On the other hand, for $n>2$ eq. (\ref{latef}) has the general form
\begin{equation}
    f(R) = R \; + \; \alpha \; R^{-k} \; ; \;\;\;  k > 0 , \alpha > 0
\end{equation}
This form for $f(R)$ was proposed by \cite{Carroll:2003wy} and \cite{Capozziello:2003gx} to give rise to late time acceleration. It was also studied in detail in \cite{Amendola:2006we}, where it was shown that this model does not reach a stable accelerated attractor after the matter epoch.

Friedmann equations associated with this $f(R)$ function are given from the field equation (\ref{fieldeq}) to be
\begin{equation} \label{frlat1}
    H^2= \frac{2^{n/2} n  \sqrt{3} \beta ^{n/2} \left(8 H^4+4 H^2 \Dot{H} n+H \Ddot{H} (n-2)+2 \Dot{H}^2\right)}{(n-2) \left(2 H^2+\Dot{H}\right) \left(2^{\frac{n}{2}+1} \sqrt{3} \beta ^{n/2}-3^{\frac{n}{2}+1} n \left(2 H^2+\Dot{H}\right)^{n/2}\right)}
\end{equation}
and
\begin{equation} \label{frlat2}
\begin{split}
    \frac{\Ddot{a}}{a} &= \left(\frac{-2^{\frac{n}{2}-1} \sqrt{3} n \beta ^{n/2}}{2(n-2) \left(2 H^2+\dot{H}\right)^2 \left(\sqrt{3} 2^{\frac{n}{2}+1} \beta ^{n/2}-3^{\frac{n}{2}+1} n \left(2 H^2+\dot{H}\right)^{n/2}\right)}\right) \Bigg[ -64 H^6 \\
    & -16 H^4 \dot{H} (n+4) -20 H^3 \ddot{H} (n-2) -4 H^2 \left(\dddot{H} (n-2)+2 \dot{H}^2 \left(-2 n^2+3 n+8\right)\right)  \\
    & +2 H \ddot{H} \dot{H} \left(4 n^2-5 n-6\right) -2 \dddot{H} \dot{H} n+4 \dddot{H} \dot{H}+\ddot{H}^2 \left(n^2-4\right)-8 \dot{H}^3 n+8 \dot{H}^3\Bigg]
\end{split}
\end{equation}

The effective energy density and pressure associated with this gravity function are then given by
\begin{equation} \label{refflat}
    \rho_eff=\frac{2^{n/2} 3 n  \sqrt{3} \beta ^{n/2} \left(8 H^4+4 H^2 \Dot{H} n+H \Ddot{H} (n-2)+2 \Dot{H}^2\right)}{(n-2) \left(2 H^2+\Dot{H}\right) \left(2^{\frac{n}{2}+1} \sqrt{3} \beta ^{n/2}-3^{\frac{n}{2}+1} n \left(2 H^2+\Dot{H}\right)^{n/2}\right)}
\end{equation}
   \begin{equation} \label{peflat}
       \begin{split}
         P_eff&=\left(\frac{\sqrt{3} 2^{\frac{n}{2}-1} n \beta ^{n/2}}{(n-2) \left(2 H^2+\Dot{H}\right)^2 \left(\sqrt{3} 2^{\frac{n}{2}+1} \beta ^{n/2}-3^{\frac{n}{2}+1} n \left(2 H^2+\Dot{H}\right)^{n/2}\right)}\right) \times \\
    &\Big[-16 H^4 \Dot{H} (2 n+5) 
     -24 H^3 \Ddot{H} (n-2)-4 H^2 \left(\dddot{H} (n-2)+2 \Dot{H}^2 \left(-2 n^2+4 n+9\right)\right)   \\
    &-96 H^6 +4 H \Ddot{H} \Dot{H} \left(2 n^2-3 n-2\right) -2 \dddot{H} \Dot{H} n+4 \dddot{H} \Dot{H}+\Ddot{H}^2 \left(n^2-4\right)-8 \Dot{H}^3 n+4 \Dot{H}^3\Big] 
       \end{split}
   \end{equation}
While the equation of state parameter will be
\begin{equation}  \label{eoslt}
   \omega_{eff}=\frac{-2 (n-2) \left(2 H^2+\Dot{H}\right) \left(8 H^2 \Dot{H}+6 H \Ddot{H}+\dddot{H}+4 \Dot{H}^2\right)-12 \left(2 H^2+\Dot{H}\right)^3+(n-2) (n+2) (4 H \Dot{H}+\Ddot{H})^2}{6 \left(2 H^2+\Dot{H}\right) \left(8 H^4+4 H^2 \Dot{H} n+H \Ddot{H} (n-2)+2 \Dot{H}^2\right)}  
\end{equation}
For the continuity equation $\nabla^\mu T^{eff}_{\mu \nu}=0$, we have 
\begin{equation}   \label{cons1}
  \begin{split}
      \Dot{\rho_{eff}} &= \left( \frac{-3\left(2^{\frac{n}{2}-1}\right) n\;H\; \beta ^{n/2}}{\left(2 H^2+\dot{H}\right)^2 \left(2^{\frac{n}{2}+1} \beta ^{n/2}-3^{\frac{n+1}{2}} n \left(2 H^2+\dot{H}\right)^{n/2}\right)}\right) \Bigg[16 H^4 \dot{H}-12 H^3 \ddot{H} \\
    & -4 H^2 \left(\dddot{H}-2 \dot{H}^2 (2 n+3)\right)+2 H \ddot{H} \dot{H} (4 n+5)-2 \dot{H} \left(\dddot{H}+4 \dot{H}^2\right)+\ddot{H}^2 (n+2)\Bigg]
  \end{split}
\end{equation}
and
\begin{equation} \label{cons2}
  \begin{split}
      \rho_{eff} \;+\;P_{eff} &= \left(\frac{\left(2^{\frac{n}{2}-1}\right) n\; \beta ^{n/2}}{\left(2 H^2+\dot{H}\right)^2 \left(2^{\frac{n}{2}+1} \beta ^{n/2}-3^{\frac{n+1}{2}} n \left(2 H^2+\dot{H}\right)^{n/2}\right)} \right) \Bigg[ 16 H^4 \dot{H}-12 H^3 \ddot{H} \\
    &-4 H^2 \left(\dddot{H}-2 \dot{H}^2 (2 n+3)\right)+2 H \ddot{H} \dot{H} (4 n+5)-2 \dot{H} \left(\dddot{H}+4 \dot{H}^2\right)+\ddot{H}^2 (n+2)\Bigg]
  \end{split}
\end{equation}
Which verifies the continuity equation, which can be written in the form
\begin{equation}
    \Dot{\rho_{eff}} \;+\; 3\; H\; \left(\rho_{eff} \;+\;P_{eff}\;\right) \;=\;0
\end{equation}

%%%%%%%%%%%%%%%%%%%%%%%%%%%%%%%%%%%%%%%%%%%%%%%%%%%%%%%%%%%

\subsection{Early Universe}
For the early universe, we assume that $\Tilde{\rho} \rightarrow \infty$, so that 
\begin{equation} \label{lrg}
    \sqrt{\frac{\Tilde{\rho}^n+1}{\Tilde{\rho}^n}} \approx 1
\end{equation}
In this case, eq. (\ref{phi1}) reads
\begin{equation} \label{erlyphi}
    \varphi(\Tilde{\rho}) \;\approx\; \pm \; \frac{2}{n \sqrt{3}} \left({\rm arcsinh}{\left(\sqrt{\Tilde{\rho}^n}\right)}-1\right) \;+\; C
\end{equation}
This can be inverted to give
\begin{equation} \label{erlyrho}
    \Tilde{\rho} \;\approx\; {\rm sinh}^\frac{2}{n} \left(1 \;\pm\; \frac{n\sqrt{3}}{2} \varphi \right)
\end{equation}
While from (\ref{R2}) $R \approx \Tilde{\beta}\;\Tilde{\rho} \;=\; \rho$, so that relation (\ref{erlyphi}) gives
\begin{equation} \label{erlyR}
    \varphi(R) \;\approx\; \pm\; \frac{2}{n \sqrt{3}} \left({\rm arcsinh} \sqrt{\left(\frac{R}{\Tilde{\beta}}\right)^n}-1\right) \;+\; C
\end{equation}

Treating (\ref{V}) for large $\Tilde{\rho}$, we get
\begin{equation}   \label{lrgV}
    V(\Tilde{\rho}) \approx \frac{1}{2} \Tilde{\beta}\;\Tilde{\rho}
\end{equation}
Using (\ref{erlyrho}) we get
\begin{equation} \label{erlyV}
    V(\varphi) \;=\; \frac{1}{2} \Tilde{\beta}\; {\rm sinh}^\frac{2}{n} \left(1 \;\pm\; 
 \frac{n\sqrt{3}}{2} \varphi\right)
\end{equation}

Now using (\ref{erlyR}) in (\ref{phid}) and integrate, we get 
\begin{equation} \label{erlyf}
\begin{split}
    f(R) & = R \pm \frac{2}{n \sqrt{3}}\left( {\rm arcsinh}\sqrt{\left(\frac{R}{\Tilde{\beta}}\right)^n} - 1 \right) R  \\ 
    & \hspace{2.1cm}  - \frac{n}{(n+2)} \; \sqrt{\left(\frac{R}{\gamma \Tilde{\beta}}\right)^{n}} \; {_2}F_1\left(\frac{1}{2},\frac{1}{2}+\frac{1}{n};\frac{3}{2}+\frac{1}{n};-\left(\frac{R}{\gamma \Tilde{\beta}}\right)^n \right) R \;+\; C_1
  \end{split}  
\end{equation} 
The Gaussian hypergeometric function $_2F_1(a,b;c;z)$ decays to zero for large negative values of z. Since we deal with a function describing the early times, we can ignore this part and consider
\begin{equation} \label{erlyft}
    f(R) = R \pm \frac{2}{n \sqrt{3}}\left({\rm arcsinh}\sqrt{\left(\frac{R}{\Tilde{\beta}}\right)^n} - 1\right)\;R\;+\;C_1
\end{equation}
Again the viability of $f(R)$ requires disregarding the $-ve$ sign function. Accordingly, the viable $f(R)$ is given by
\begin{equation} \label{erlyv}
    f(R) = R + \frac{2}{n \sqrt{3}}\left({\rm arcsinh}\sqrt{\left(\frac{R}{\Tilde{\beta}}\right)^n} - 1\right)\;R\;+\;C_1
\end{equation}
A quick test for the function in (\ref{erlyv}) reveals that the $\rm arcsinh$-term tends to be a logarithmic function as $R$ increases, which is a slowly varying function, so that $f(R) \sim C\; R$ for large $R$. This may permit the function to describe the inflationary era.

Friedmann equations for this function are given by
\begin{equation} \label{frerl1}
    H^2= \frac{3n \sqrt{\Tilde{R}^n} \left(2 \left(2 H^2+\dot{H}\right)^2 \left(\Tilde{R}^n+1\right)- H (4 H \dot{H}+\ddot{H}) \left(2 \left(\Tilde{R}^n+1\right)+n\right)\right)}{R (\Tilde{R}^n+1)^{3/2} 
       \left(\frac{n \sqrt{\Tilde{R}^n}}{ \sqrt{\Tilde{R}^n+1}} +2 \sinh ^{-1}\left(\sqrt{\Tilde{R}^n}\right)  + n \sqrt{3} -2 \right)}
\end{equation}
and
\begin{equation} \label{frerl2}
\begin{split}
    \frac{\Ddot{a}}{a} &= \left( \frac{n \Tilde{R}^n}{6 R^2 \sqrt{\Tilde{R}^n} \left(\Tilde{R}^n+1\right)^2 \left(n  \sqrt{\Tilde{R}^n} +\sqrt{3} n \sqrt{\Tilde{R}^n+1} -2 \sqrt{\Tilde{R}^n+1} +2 \sqrt{\Tilde{R}^n+1} \sinh ^{-1}\left(\sqrt{\Tilde{R}^n}\right)\right)}\right) \times \\
    & \Bigg[36 H^2 \dot{H} \left[12 \dot{H} \left(n^2 (2 \Tilde{R}^n-1)+4 (\Tilde{R}^n+1)^2\right)-R \left(\Tilde{R}^n+1\right) \left(2 \Tilde{R}^n+n+2\right)\right] \\
    & +9 H \ddot{H} \left[24 \dot{H} \left(n^2 \left(2 \Tilde{R}^n-1\right)+4 \left(\Tilde{R}^n+1\right)^2\right)-5 R \left(\Tilde{R}^n+1\right) \left(2 \Tilde{R}^n+n+2\right)\right]  \\
    & +R \left(\Tilde{R}^n+1\right) \left[-9 \dddot{H} \left(2 \Tilde{R}^n+n+2\right)-36 \dot{H}^2 \left(2 \Tilde{R}^n+n+2\right)+R^2 \left(\Tilde{R}^n+1\right)\right] \\
    &+27 \ddot{H}^2 \left[n^2 \left(2 \Tilde{R}^n-1\right)+4 \left(\Tilde{R}^n+1\right)^2\right]\Bigg]
\end{split}
\end{equation}

Where $\Tilde{R} = R/\beta$. The effective energy density and pressure are then given by
\begin{equation} \label{rear}
   \rho_{eff}\;=\; \frac{9 n \sqrt{\Tilde{R}^n} \left(2 \left(2 H^2+\dot{H}\right)^2 \left(\Tilde{R}^n+1\right)- H (4 H \dot{H}+\ddot{H}) \left(2 \left(\Tilde{R}^n+1\right)+n\right)\right)}{R (\Tilde{R}^n+1)^{3/2} 
       \left(\frac{n \sqrt{\Tilde{R}^n}}{ \sqrt{\Tilde{R}^n+1}} +2 \sinh ^{-1}\left(\sqrt{\Tilde{R}^n}\right)  + n \sqrt{3} -2 \right)}
\end{equation}
and
\begin{equation} \label{pear}
 \begin{split}
   P_{eff}\;=& -\left(\frac{n\;\sqrt{\Tilde{R}^n}}{4\;R^2 (\Tilde{R}^n+1)^2 \left(n \sqrt{3} \sqrt{\Tilde{R}^n+1} -2 \sqrt{\Tilde{R}^n+1}+2 \sqrt{\Tilde{R}^n+1} \sinh ^{-1}\left(\sqrt{\Tilde{R}^n}\right)+n \;\sqrt{\Tilde{R}^n}\right)}\right) \times  \\
   & \Bigg[36 (4 H \dot{H}+\ddot{H})^2 
     \left(4 (\Tilde{R}^n+1)^2+2\; n^2 \Tilde{R}^n \right)+2 (\Tilde{R}^n+1)^2 R^3  \\
     & \;-12\; R (\Tilde{R}^n+1) \left[\left(4 H \ddot{H}+\dddot{H}+4 \dot{H}^2\right)+2 H (4 H \dot{H}+\ddot{H})\right]\left(2\Tilde{R}^n+n+2\right) \Bigg] 
 \end{split}
\end{equation}
The equation of state parameter will be
\begin{equation} \label{eosear}
    \begin{split}
        \omega_{eff}\;=&-\left(\frac{1}{R \left(\Tilde{R}^n+1\right) \left(R^2 (\Tilde{R}^n+1) -18 H \left(2 (\Tilde{R}^n+1)+n\right) \left(4 H \dot{H}+\ddot{H}\right)\right)}\right) \times \\
        & \Bigg[ 18\left(4 H \dot{H}+\ddot{H}\right)^2 \left(4 \left(\Tilde{R}^n+1\right)^2+n^2 \left(2\; \Tilde{R}^n-1\right)\right) \\
        & + R^3 (\Tilde{R}^n+1)^2 -\;3\;R \left(\Tilde{R}^n+1\right) \left(2 (\Tilde{R}^n+1)+n\right) \left(8 H^2 \dot{H}+6 H \ddot{H}+\dddot{H}+4 \dot{H}^2\right) \Bigg]
    \end{split}
\end{equation}
For verification of the continuity equation, we have
\begin{equation} \label{conear1}
    \begin{split}
        \Dot{\rho}=& \left(\frac{27\; n\; H  \sqrt{\Tilde{R}^n}}{\Tilde{R}^2\; (\Tilde{R}^n+1)^2  \left(n \sqrt{ \Tilde{R}^n}+ n \sqrt{3} \sqrt{\Tilde{R}^n+1} -2 \sqrt{\Tilde{R}^n+1}+2 \sqrt{\Tilde{R}^n+1} \sinh ^{-1}\left(\sqrt{\Tilde{R}^n}\right)\right)}\right) \times \\
        & \Bigg[16 (\Tilde{R}^n+1) H^4 \dot{H} \left(2 (\Tilde{R}^n+1)+n\right)-12 (\Tilde{R}^n+1) H^3 \ddot{H} \left(2 (\Tilde{R}^n+1)+n\right)   \\
        & -4 H^2 \Bigg((\Tilde{R}^n+1) \dddot{H} (2 (\Tilde{R}^n+1)+n)- 2 \dot{H}^2 \left(-(\Tilde{R}^n+1) n+6 (\Tilde{R}^n+1)^2+2\;n^2 (2 \Tilde{R}^n-1)\right)\Bigg) \\
        & +\ddot{H}^2 \left(4 (\Tilde{R}^n+1)^2+n^2 (2\Tilde{R}^n-1)\right)+2 H \ddot{H} \dot{H} \left(-3 (\Tilde{R}^n+1) n+10 (\Tilde{R}^n+1)^2+4\;n^2 (2 \Tilde{R}^n-1)\right) \\
        & -2 (\Tilde{R}^n+1) \dot{H} (2 (\Tilde{R}^n+1)+n) (\dddot{H}+4 \dot{H}^2)\Bigg]
    \end{split}
\end{equation}
and
\begin{equation} \label{conear2}
    \begin{split}
        \rho_{eff}+P_{eff}\;=& -\left(\frac{9\; n\;  \sqrt{\Tilde{R}^n}}{\Tilde{R}^2\; (\Tilde{R}^n+1)^2  \left(n \sqrt{ \Tilde{R}^n}+ n \sqrt{3} \sqrt{\Tilde{R}^n+1} -2 \sqrt{\Tilde{R}^n+1}+2 \sqrt{\Tilde{R}^n+1} \sinh ^{-1}\left(\sqrt{\Tilde{R}^n}\right)\right)}\right) \times \\
        & \Bigg[16 (\Tilde{R}^n+1) H^4 \dot{H} \left(2 (\Tilde{R}^n+1)+n\right)-12 (\Tilde{R}^n+1) H^3 \ddot{H} \left(2 (\Tilde{R}^n+1)+n\right)   \\
        & -4 H^2 \Bigg((\Tilde{R}^n+1) \dddot{H} (2 (\Tilde{R}^n+1)+n)- 2 \dot{H}^2 \left(-(\Tilde{R}^n+1) n+6 (\Tilde{R}^n+1)^2+2\;n^2 (2 \Tilde{R}^n-1)\right)\Bigg) \\
        & +\ddot{H}^2 \left(4 (\Tilde{R}^n+1)^2+n^2 (2\Tilde{R}^n-1)\right)+2 H \ddot{H} \dot{H} \left(-3 (\Tilde{R}^n+1) n+10 (\Tilde{R}^n+1)^2+4\;n^2 (2 \Tilde{R}^n-1)\right) \\
        & -2 (\Tilde{R}^n+1) \dot{H} (2 (\Tilde{R}^n+1)+n) (\dddot{H}+4 \dot{H}^2)\Bigg]
    \end{split}
\end{equation}
Which verifies the continuity equation.

%%%%%%%%%%%%%%%%%%%%%%%%%%%%%%%%%%%%%%

\section{Inflationary behavior}
\subsection{Inflation Observables}
\begin{figure} %[h!]
    \centering
    \includegraphics[width=0.5 \textwidth]{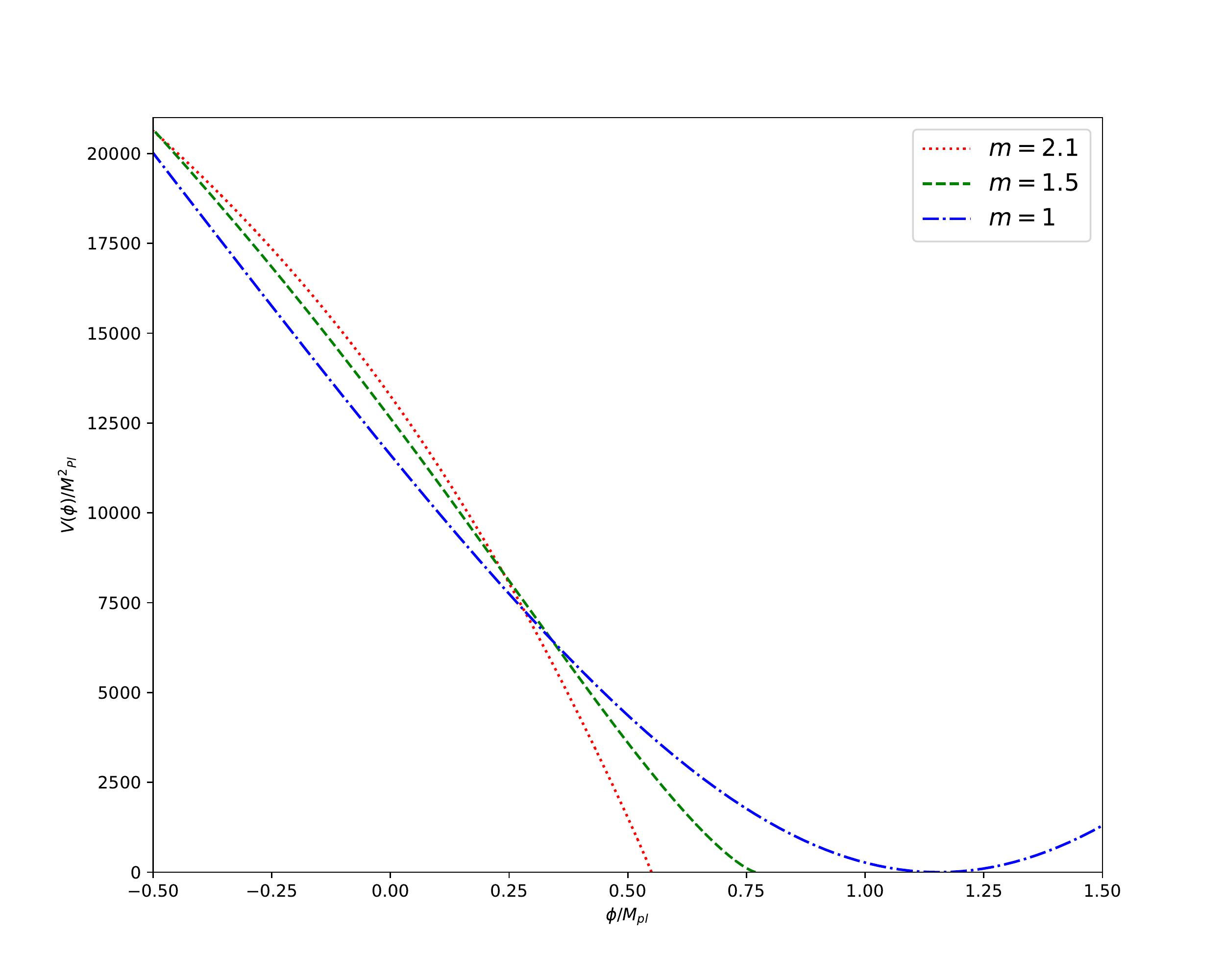}
    \caption{\footnotesize Potential function at different values of $n$.}
    \label{pot}
\end{figure}

let's now study the ability of our constructed $f(R)$ function, to describe the inflationary era. our potential function as derived in (\ref{erlyV}) is
\begin{equation} 
    V(\varphi) \;=\; \frac{1}{2} \Tilde{\beta}\; {\rm sinh}^\frac{2}{n} \left(1 \;\pm\; 
 \frac{n\sqrt{3}}{2} \varphi\right) \nonumber
\end{equation}
This potential function is displayed in Fig. \ref{pot}. Now we will test this potential for slaw roll inflation. The slow roll parameters due to a given potential are given by

 \begin{equation}
 \begin{split}
     \epsilon_V & = \frac{M_P^2}{2} \left(\frac{V_{,\varphi}}{V}\right)^2 \\
     \eta_V & = M_P^2 \left[ \frac{V_{,\varphi \varphi}}{V} - \frac{1}{2} \left(\frac{V_{,\varphi}}{V}\right)^2 \right]
 \end{split}
 \end{equation}

 Using our potential in eq. (\ref{erlyV}), we get
 \begin{equation} \label{eps}
     \epsilon_V = \frac{3}{2 \left[(1-\frac{n \sqrt{3}}{2} \varphi )^2+1\right] \left[\sinh ^{-1}(1-\frac{n \sqrt{3}}{2} \varphi )\right]^2}\;M_P^2
 \end{equation}
 and 
 \begin{equation}
     \eta_V = \frac{3}{2\left[\sinh ^{-1}(1-\frac{n \sqrt{3}}{2} \varphi )\right]^2} \left(\frac{ \left(2-n\right)}{(1-\frac{n \sqrt{3}}{2} \varphi )^2+1}-\frac{n (1-\frac{n \sqrt{3}}{2} \varphi ) \sinh ^{-1}(1-\frac{n \sqrt{3}}{2} \varphi )}{\left[(1-\frac{n \sqrt{3}}{2} \varphi )^2+1\right]^{3/2}}\right) \;M_P^2
 \end{equation}

 At the end of inflation, we have $\epsilon_V(\varphi_e) \simeq 1$. Using this in eq. (\ref{eps}) and solve numerically for $\varphi$, we get
 \begin{equation}
     \varphi_e = \frac{0.026834}{n\sqrt{3}} \; M_P
 \end{equation}
 On the other hand, the duration of the inflation is measured by the number of e-folds, $N$, given by
 \begin{equation}
     N = \frac{1}{M_P^2} \int_{\varphi_e}^{\varphi_N}{\frac{V(\varphi)}{V_{,\varphi}(\varphi)}} d \varphi
 \end{equation}
Using eq. (\ref{erlyV}), we get
\begin{equation}
 \begin{split}
    N & = \frac{1}{M_P^2} \left[ \frac{1}{3 n \sqrt{2}} \left\{  \left(\frac{n \sqrt{3} \varphi}{2}-1\right)^2  - \left(\sinh ^{-1}\left(1-\frac{n \sqrt{3}}{2}  \varphi \right)\right)^2 \right. \right. \\
    &  \;\;\;\;\;\;\; \left. \left. + \; 2 \left(\frac{n \sqrt{3} \varphi}{2}-1\right) \sqrt{2 + \frac{n \sqrt{3}  \varphi}{2} \left(\frac{n \sqrt{3} \varphi}{2}-2\right)} \; \sinh ^{-1}\left(1-\frac{n \sqrt{3}}{2}  \varphi\right) 
 \right\} \right]_{\varphi_e}^{\varphi_N}
 \end{split}
\end{equation}

 This gives the field $\varphi_N$ for a given number of e-folds, $N$, before the end of inflation. The tensor-to-scalar ratio $r$ and the scalar spectral index $n_s$ can now be calculated for a given $N$ from the relations
\begin{equation}
 \begin{split}
     r & = 16 \; \epsilon_V(\varphi_N) \\
     n_s & = 1 \; -\; 4\; \epsilon_V(\varphi_N) + 2\; \eta_V(\varphi_N)
 \end{split}
 \end{equation}
%%%%%%%%%%%%%%%%%%%%%%%%%%%%%%%%%%%%%5

\subsection{Observational Constraints}
Let's test the viability of the potential (\ref{erlyV}), and in consequence the ability of our constructed function to describe inflation. In particular, recent data from Planck 2018 \cite{Planck:2018jri} puts an upper bound on the tensor-to-scalar ratio, $r_{0.002} < 0.1$ up to $95\%$ CL. This is further tightened by combining with the BICEP2/Keck Array BK15 data to be $r_{0.002} < 0.056$. The spectral index of the scalar perturbation is determined by the Planck temperature data in combination with the EE measurements at low multipoles to be $n_s = 0.9626 \pm 0.0057$ at $68 \%$ CL.

\begin{figure} [h!]
    \centering
    \includegraphics[width=0.5 \textwidth]{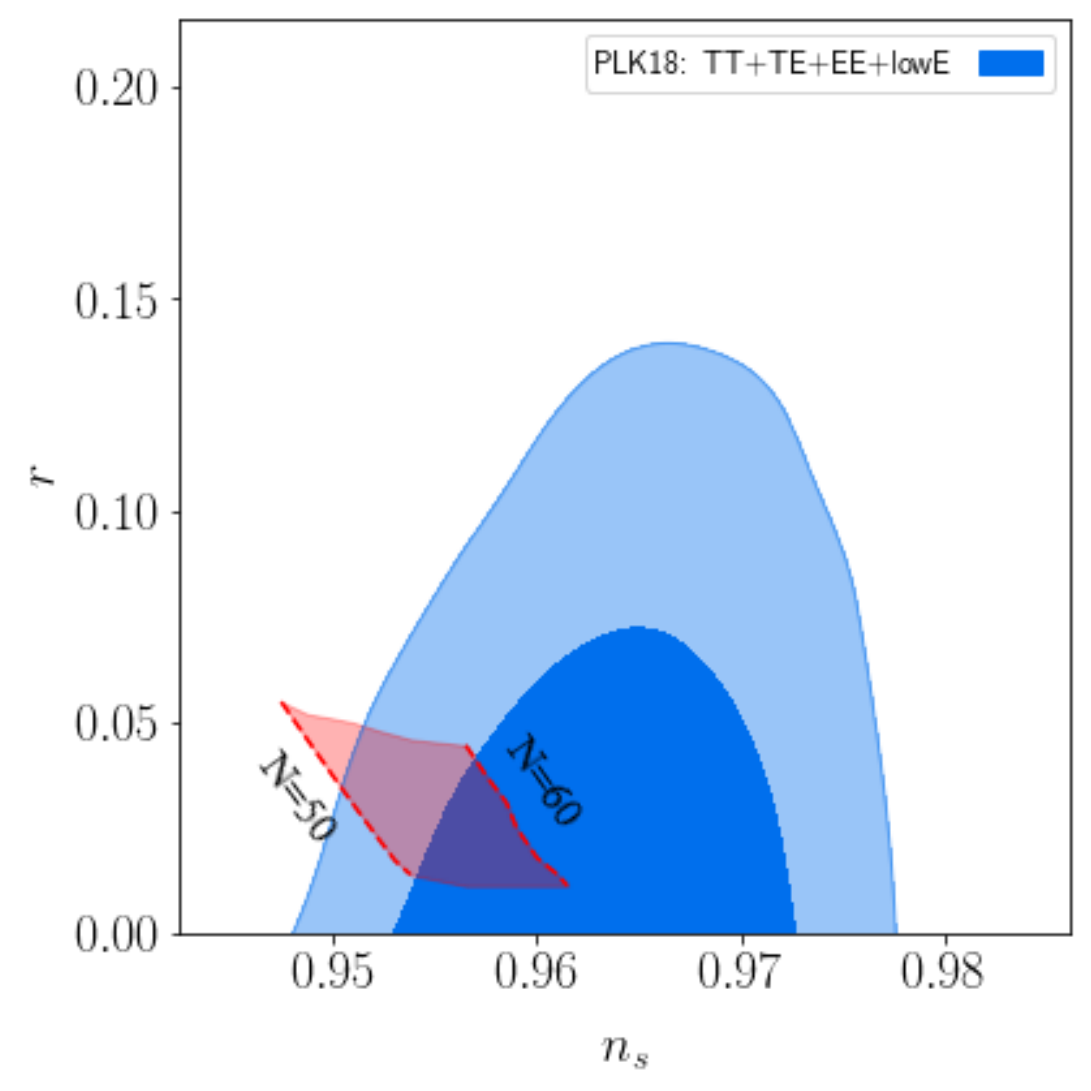}
    \caption{\footnotesize $n_s - r$ contours for Planck 2018 data \cite{Planck:2018jri}. The red patch is the result of our potential due to the constructed $f(R)$ function for $1.8 \le n \le 6$.}
    \label{obs(z)}
\end{figure}

In Fig. \ref{obs(z)} we present Planck 2018 contours for the $(n_s, r)$ plane. On top of that, we display the area for these two parameters based on the potential (\ref{erlyV}) for the model parameter $1.8 \le n \le 6$. We can see that our results are consistent with those of Planck 2018 TT+TE+EE+lowE data. Particularly, the tensor-to-scalar ratio always satisfies the criteria of Planck even for the tightened result $r < 0.056$. The scalar spectral index, on the other hand, depends on the model parameter $n$ and is consistent with the Planck data for $n > 2$.

%%%%%%%%%%%%%%%%%%%%%%%%%%%%%%%%%%%%%%

\section{Conclusions}
In this work, we exploit the connection between the scalar-tensor theory and the $f(R)$ gravity to reconstruct the new cosmological UDF model, proposed in our previous work, in the framework of $f(R)$ gravity. We derived the field equation of the model, whence constructed its $f(R)$ function and discussed its viability. We also studied the $f(R)$ functions and the scalar field potentials in the asymptotically de Sitter spacetime for the early and late times universe.

We then tested the ability of our constructed $f(R)$ function to describe the early time inflation. The scalar field potential for the early time is used to derive the slow roll inflation parameters. Tensor-to-scalar ratio $r$ and the scalar spectral index $n_s$ were calculated for different values of the model parameter $n$ and compared to recent observations from Planck 2018 data. We found that our results for the model parameter $n > 2$ agree with Planck-2018 TT+TE+EE+lowE data.

\section*{Acknowledgments}
The author would like to thank W. El Hanafy and M. Hashim for their valuable discussions.

\bibliographystyle{unsrt}%amsplain}
\bibliography{fofr}
\end{document}